\documentclass[rnote]{aa} 
\usepackage{subfigure}
\usepackage{epsfig}
\usepackage{lscape}
\usepackage{rotating}
\begin{document}

   \title{The molecular gas content of ULIRG type 2 quasars at $z<$1\thanks{Based on observations carried out with the IRAM 30m radiotelescope.}}

     \author{M. I. Rodr\'\i guez\inst{1}\and M. Villar-Mart\'\i n\inst{2}\and B. Emonts\inst{2}\and A. Humphrey\inst{3} \\    
  G. Drouart\inst{4},   S. Garc\'\i a Burillo\inst{5}, M. P\'erez Torres\inst{1}}
   \institute{Instituto de Astrof\'{\i}sica de Andaluc\'{\i}a  (CSIC), Glorieta de la Astronom\'{\i}a s/n,  Granada, Spain, \email{mrm@iaa.es}\and
Centro de Astrobiolog\'\i a (INTA-CSIC), Carretera de Ajalvir, km 4, 28850 Torrej\'on de Ardoz, Madrid, Spain\and 
Centro de Astrofisica, Universidade do Porto, Rua das Estrelas, 4150-762 Porto, Portugal\and
Dept. of Earth and Space Science, Onsala Space Observatory, Chalmers U. of Technology, SE-43992 Onsala, Sweden \and
Observatorio Astron\'omico Nacional (OAN)-Observatorio de Madrid, Alfonso XII, 3, 28014, Madrid, Spain    }

   \date{November 2013}

\abstract{
We present new results of CO(1-0) spectroscopic observations of 4 SDSS type 2 
quasars (QSO2) at z$\sim$0.3, observed with the 30m IRAM telescope. The QSO2 
have infrared luminosities in the ULIRG (UltraLuminous Infrared Galaxies)  regime. 
We confirm the  CO(1-0) detection in one of our 4 QSO2, SDSS J1543-00, with $L'_{CO}$ and $M_{H_2}$  
(1.2$\pm$0.2) $\times$10$^{10}$ K km s$^{-1}$ pc$^2$ and (9.4$\pm$1.4)$\times$10$^9$ 
M$_{\odot}$, respectively. The CO(1-0) line has $FWHM=$575$\pm$102 km s$^{-1}$. No 
CO(1-0) emission is detected in SDSS J0903+02, SDSS J1337-01, SDSS J1520-01 
above 3 sigma, yielding upper limits on $M(H_2)\sim$ 9.6, 4.3 and 5.1 
$\times$10$^9$ M$_{\odot}$ respectively.  Together with CO measurements of
 9  QSO2 at $z\sim$0.3-1.0 from the ULIRG sample  by Combes et al. (2011, 2013), 
 we expand previous studies of the molecular gas content of intermediate $z$ QSO2 into 
 the ULIRG regime. We discuss the location of the 13 ULIRG QSO2 at $z<$1 with available $L'_{CO}$ 
 measurements in the $L'_{CO}$ vs. $z$ and $L'_{CO}$ vs. $L_{FIR}$ diagrams, in comparison 
 with other QSO1 and ULIRG star forming samples. 
}


\maketitle

\section{Introduction}
Following the discovery in large numbers of type 2 quasars (QSO2) during the last 
decade, an intensive follow up has been performed at different wavelengths: X-ray, radio, 
infrared and optical (e.g., Szokoly et al. 2004, Lacy et al. 2007, Mart\'{\i}nez-Sansigre 
et al. 2006,  Zakamska et al. 2003). In spite of this, the molecular gas 
content of this class of objects has  been very scarcely studied.
  
The study of the molecular gas content in these objects is crucial to better understand the 
conditions required to trigger both the nuclear and star formation activities in the most 
luminous active galaxies (AGN), since this gaseous phase can provide large amounts of 
fuel to form stars  and feed the nuclear black hole.  It can also provide information 
about the relation between QSO2 and other systems such as  type 1 
quasars (QSO1), luminous (LIRG, 10$^{11}$ L$_{\odot} \le L_{IR}< $10$^{12}$ L$_{\odot}$) and ultra-luminous  (ULIRG, $L_{IR}\ge$10$^{12}$ L$_{\odot}$) infrared galaxies,
where $L_{IR}$ is the total infrared luminosity ($\sim$8-1000 $\mu$m range). However, only 
few molecular gas studies of QSO2 have been carried out, and frequently focussed on $z>$2 objects
(see Villar Mart\'\i n et al. 2013 (VM13 hereafter) and references therein).

 \section{QSO2 sample at $z<$1}

There are $L'_{CO}$ measurements  for 29 QSO2 at $z<$1 published in three different works. 15  CO(1-0) confirmed detections are reported
(i.e. 52\% detection rate) which  imply molecular gas masses M(H$_2$) = $\alpha\times$ $L'_{CO}$  in the range 
(0.5-15)$\times$10$^9$ M$_{\odot}$\footnote{For coherence with other works, we  assume $\alpha$=0.8 M$_{\odot}$~(K km s$^{-1}$ ~pc$^2$)$^{-1}$ (Downes \& Solomon 1998). Recent results suggest $\alpha$=0.6$\pm$0.2 (Papadopoulos et al. 2012).}.

Krips, Neri \& Cox (2012,  KNC12 hereafter)  investigated 
the molecular gas content of 10 QSO2 at z$\sim$0.1-0.4 based on observations performed with the IRAM Plateau de Bure Interferometer (PdBI).  
 All but two objects (selected from Zakamska et al. 2003), are from the original sample
of 24 $\mu$m selected galaxies observed with the Spitzer infrared
spectrograph for the 5 millijanksy Unbiased Spitzer Extragalactic
Survey (5MUSES) (Wu et al. 2010, see also Lacy et al. 2007). KNC12 confirm the detection of CO(1-0)
in five sources and a tentative detection for a sixth. The molecular gas masses were found  in
the range of M(H$_2$)$\sim$(0.4-2.6)$\times$10$^9$ M$_{\odot}$ for the detections and
upper limits are in the range (0.4-2.2)$\times$10$^9$ M$_{\odot}$ for the  non detections. 

In VM13  we presented the results of CO(1-0) spectroscopic observations of 10  QSO2 at 
z$\sim$0.2-0.34 performed with the 30m IRAM radiotelescope and the Australia Telescope Compact 
Array. All objects were selected from the original sample of QSO2  at 0.3$\la z \la$ 0.8 identified by 
Zakamska et al. (2003) from the Sloan Digital Sky Survey  (SDSS, York et al. 2000).
5 new confirmed CO(1-0) detections were reported, with  M(H$_2$)$\sim$(5-6)$\times$10$^9$ M$_{\odot}$, and 1 tentative detection. 
Upper limits are in the range (1.6-5.0)$\times$10$^9$ M$_{\odot}$ for the  non detections.

 Most of these 20 QSO2  (17/20) have $L_{IR}<$10$^{12}$  L$_{\odot}$, i.e.,  in the LIRG regime or below. Only three have $L_{IR}\sim$10$^{12}$ L$_{\odot}$
  in the transition between the LIRG and ULIRG  regimes. 
 The implied   molecular gas masses    are  found to be consistent with QSO1 of similar
   infrared luminosities. 

Here we expand our work  into the ULIRG regime. On one hand, we  include  results of CO(1-0) observations of  4  ULIRG SDSS QSO2 
at $z\sim$0.3 (Table 1), to match in $z$ the sample by VM13. 
 
  We also include  9 objects in the $z\la$1 ULIRG samples by Combes et al. (2011, 2013; C11 and C13 hereafter).  Making use of the original SDSS spectra 
  when available or optical line luminosities published
in the literature, we find that the following ULIRG can be classified as QSO2
according to Zakamska et al. (2003) criteria: G19, G30, S1, S4, S8, S10, S12, S21, S25 (the
 nomenclature in C11, C13 is used).
These 9 QSO2  are at $z\sim$0.3-0.9.
Spectroscopic observations of different CO transitions performed with the IRAM 30m radiotelescope imply 
M(H$_2$)$\sim$(3.2-15.4)$\times$10$^9$ M$_{\odot}$ for the 5 detections and upper limits $\sim$(1.1-7.2)$\times$10$^9$ M$_{\odot}$ for 
the 4 non detections (C11, C13).

We assume
$\Omega_{\Lambda}$=0.7, $\Omega_M$=0.3, H$_0$=71 km s$^{-1}$ Mpc$^{-1}$.

\section{Observations and data reduction}
\label{sec:obs}
\begin{table*}
      \caption[]{The sample. $z_{\rm SDSS}$ is the optical redshift derived from the [O III]$\lambda$5007 line using the SDSS spectra. $D_L$  is the luminosity distance
in Mpc. $t_{exp}$  gives the total exposure time per source including calibrations. 
$\nu_{obs}$ is the observed frequency of the CO(1-0) transition and rms is the noise determined from channels with 16 MHz ($\sim$50 km s$^{-1}$) spectral resolution  in mK.}
      \tiny
    \begin{tabular}{p{0.15\linewidth}lllllllllll}
            \hline
            \noalign{\smallskip}
Object &  RA(2000) & Dec(2000) & z$_{\rm SDSS}$ & $D_L$ & $t_{exp}$ & $\nu_{obs}$ & T$_{sys}$ & rms \\
	& 		&		& 	& Mpc &	hr & GHz   & K & mK \\ \hline
            \noalign{\smallskip}
SDSS J0903+02 & 09:03:07.84  & +02:11:52.2  & 0.3290 & 1716 & 2.6 & 86.74 & 101 &0.60 \\
SDSS J1337-01  & 13:37:35.02 & -01:28:15.7   & 0.3282 & 1711 & 5.5 & 86.79 & 88 & 0.27 \\
SDSS J1520-01  & 15:20:19.75 & -01:36:11.2   & 0.3070 & 1583  & 4.4 & 88.20 & 93 & 0.39 \\					    
SDSS J1543-00  & 15:43:37.82 & -00:44:19.9   & 0.3107 & 1605  & 7.1 & 87.95 & 91 & 0.25 \\
         \noalign{\smallskip}
            \hline
         \end{tabular}
 \end{table*}

\begin{table*}
\normalsize
\caption[]{The luminosity (2) of the [OIII]$\lambda$5007 line (Zakamska et al. 2003) is given in logarithmic units
of L$_{\odot}$.
$L'_{CO}$  (3) is in units of $\times$10$^9$ K km s$^{-1}$ pc$^2$, where the upper limits correspond to 3 $\sigma$. A conversion factor $\alpha$=0.8 M$_{\odot}$~(K km s$^{-1}$ pc$^2$)$^{-1}$ has been assumed to calculate the molecular gas mass (4),  with $M(H_2)$ = $\alpha\times$ $L'_{CO}$. V$_{\rm CO - [OIII]}$ is the velocity redshift of the CO(1-0) line relative to $z_{\rm SDSS}$.
 Columns (8) and (9) give   the total infrared ($\sim$8-1000 $\mu$m) and far infrared ($\sim$40-500 $\mu$m) IRAS fluxes in mJy, while the luminosities  are shown in (10) and (11) in units of $\times$10$^{12}$ L$_{\odot}$.}
      \tiny
    \begin{tabular}{p{0.15\linewidth}lllllllllllll}
            \hline
            \noalign{\smallskip}
Object &  log[OIII]  & $L'_{CO}$ &  $M(H_2)$& FWHM$_{CO}$ &  V$_{CO-[OIII]}$ & F$_{60 \mu m}$ &  F$_{100 \mu m}$  & $L_{FIR}$  & $L_{IR}$\\
		 &  & $\times$10$^9$ &  $\times$10$^9$ M$_{\odot}$  & km s$^{-1}$ & km s$^{-1}$ & mJy	&	mJy	&  $\times$10$^{12}$ L$_{\odot}$  & $\times$10$^{12}$  L$_{\odot}$   \\ 
		 (1) & (2) & (3) & (4) & (5) & (6) & (7)  & (8) & (10) & (11) \\ \hline 
SDSS J0903+02 & 8.42   & $<$ 12.0   & $<$ 9.6 & -        & -     & 390 & 540  &  2.5 & 3.8 \\
SDSS J1337-01  & 8.72  & $<$ 5.3     & $<$ 4.3  & -   &  - &  140 & N/A & 1.4$^{+0.8}_{-0.3}$  &  2.1$^{+1.2}_{-0.45}$  \\
SDSS J1520-01  &  8.29  &  $<$ 6.8  & $<$ 5.4  & -         &  -  & N/A  & 490 & 1.2$^{+0.3}_{-0.3}$ & 1.8$^{+0.45}_{-0.4}$ \\    
SDSS J1543-00  & 8.40  & 11.7 $\pm$ 1.8   & 9.4 $\pm$ 1.4  & 575 $\pm$ 102 &    128 $\pm$ 47 &  130 & N/A & 1.1$^{+0.7}_{-0.2}$ & 1.65$^{+1.1}_{-0.3}$ \\
         \noalign{\smallskip}
            \hline
         \end{tabular}
 \end{table*}

The observations were carried out using the 30m IRAM single dish telescope at Pico Veleta, 
during two different observing runs in February and June 2013 (programme 202-12). The EMIR receiver was tuned 
to the redshifted frequencies of the CO(1-0) line (115.27 GHz rest frame), using the optical SDSS $z$  for each object (see Table 
1). The observations were performed in the wobbler-switching mode with a throw  50"  in order 
to ensure flat baselines. We observe both polarizations (H and V) using as a 
backend the WILMA autocorrelator that produced an effective total bandwidth 
of 4 GHz with a (Hanning-smoothed) velocity resolution of 16 MHz $\sim$ 50 km s$^{-1}$. The 
corresponding T$_{sys}$, total integration time and the rms corresponding for 
each source is specified in Table \ 1. The temperature scale used is in main beam temperature 
T$_{mb}$. At 3mm the telescope half-power beam width is 29$^{''}$. The main-beam 
efficiency is $\eta_{mb}$= T$^*_{A}$/T$_{mb}$ =0.81 and the conversion factor is $S$/T$^*_{A}$ = 5.9 Jy/K.

The pointing model was checked against bright, nearby calibrators for every 
source, and every 1.6 hrs for long integrations. It was found to be accurate 
within 5$"$. Calibration scans on the standard two load system were 
taken every 8 minutes.

The off-line data reduction was done with the CLASS program of the 
GILDAS software package (Guilloteau \& Forveille 1989), and 
involved only the subtraction of (flat) baselines from individual
integrations and the averaging of the total spectra (Fig.~1).

\begin{figure}[!h]
     \centering
     \subfigure
         {\label{fig:f1}}
          \includegraphics[width=.38\textwidth, angle= 0]{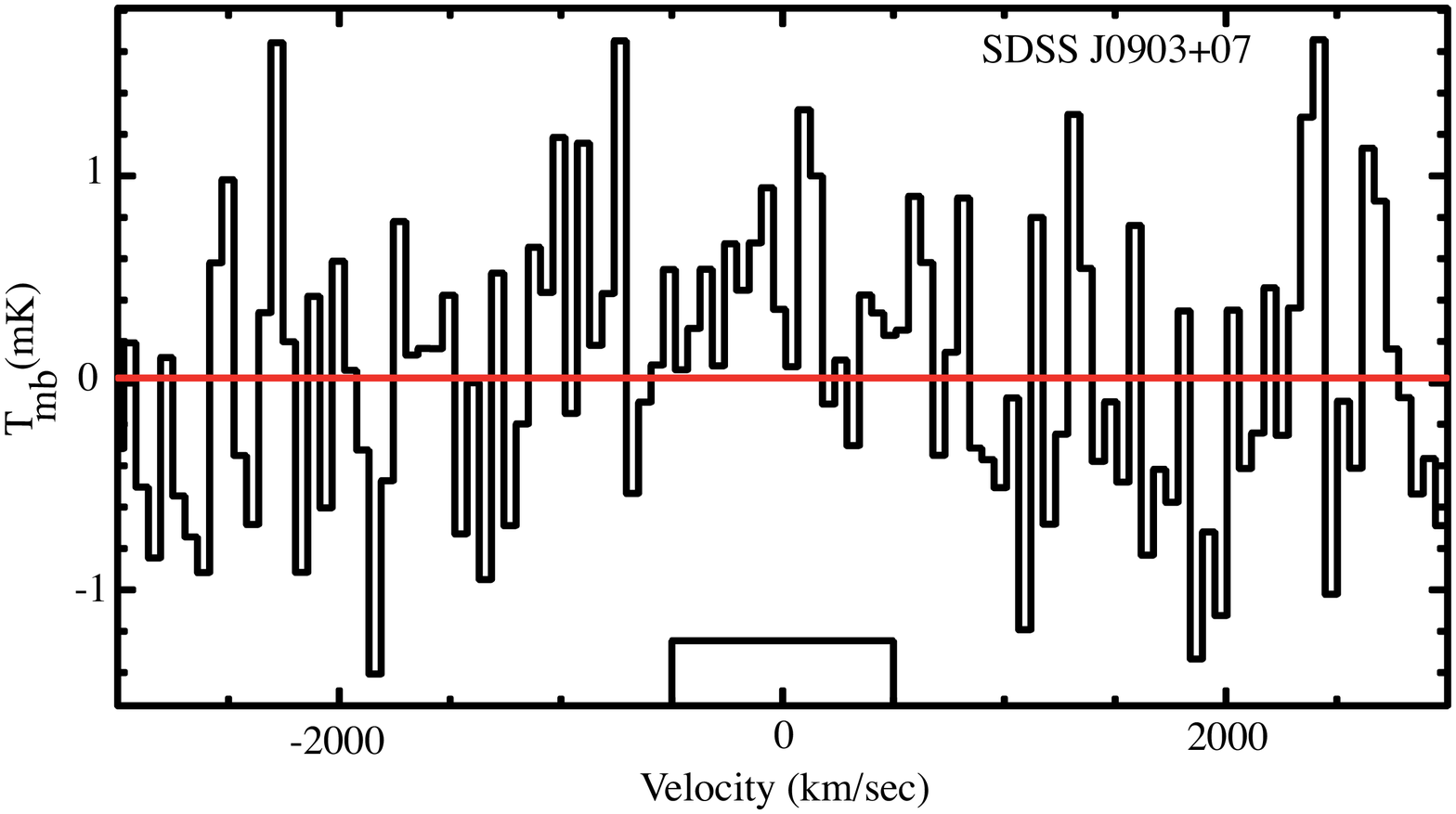}
          \hspace{-0.10in}
     \subfigure
          {\label{fig:f2}}
          \includegraphics[width=.38\textwidth, angle= 0]{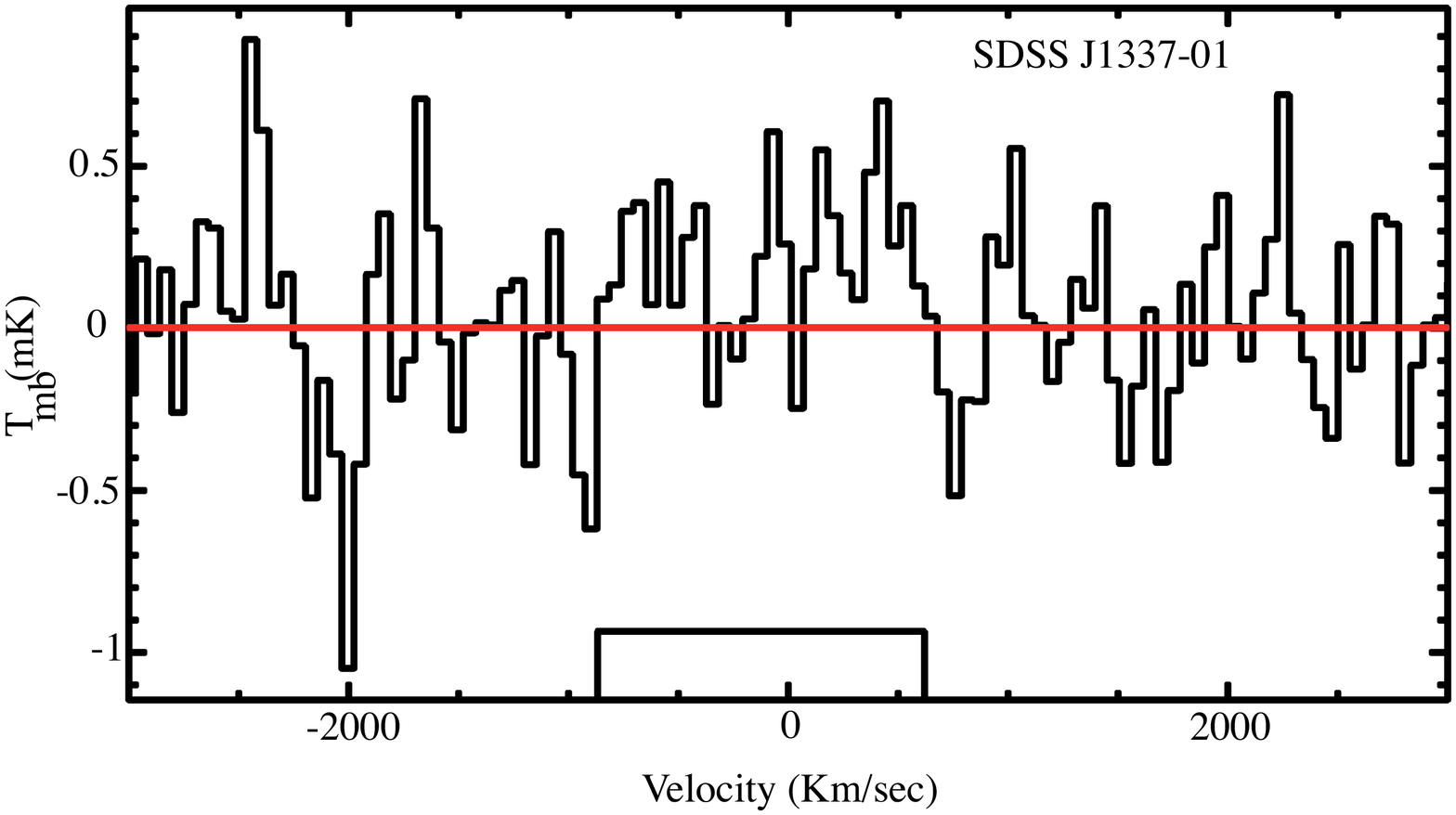}
                 \vspace{-0.1in}                
    \subfigure
         {\label{fig:f3}}
          \includegraphics[width=.38\textwidth, angle= 0]{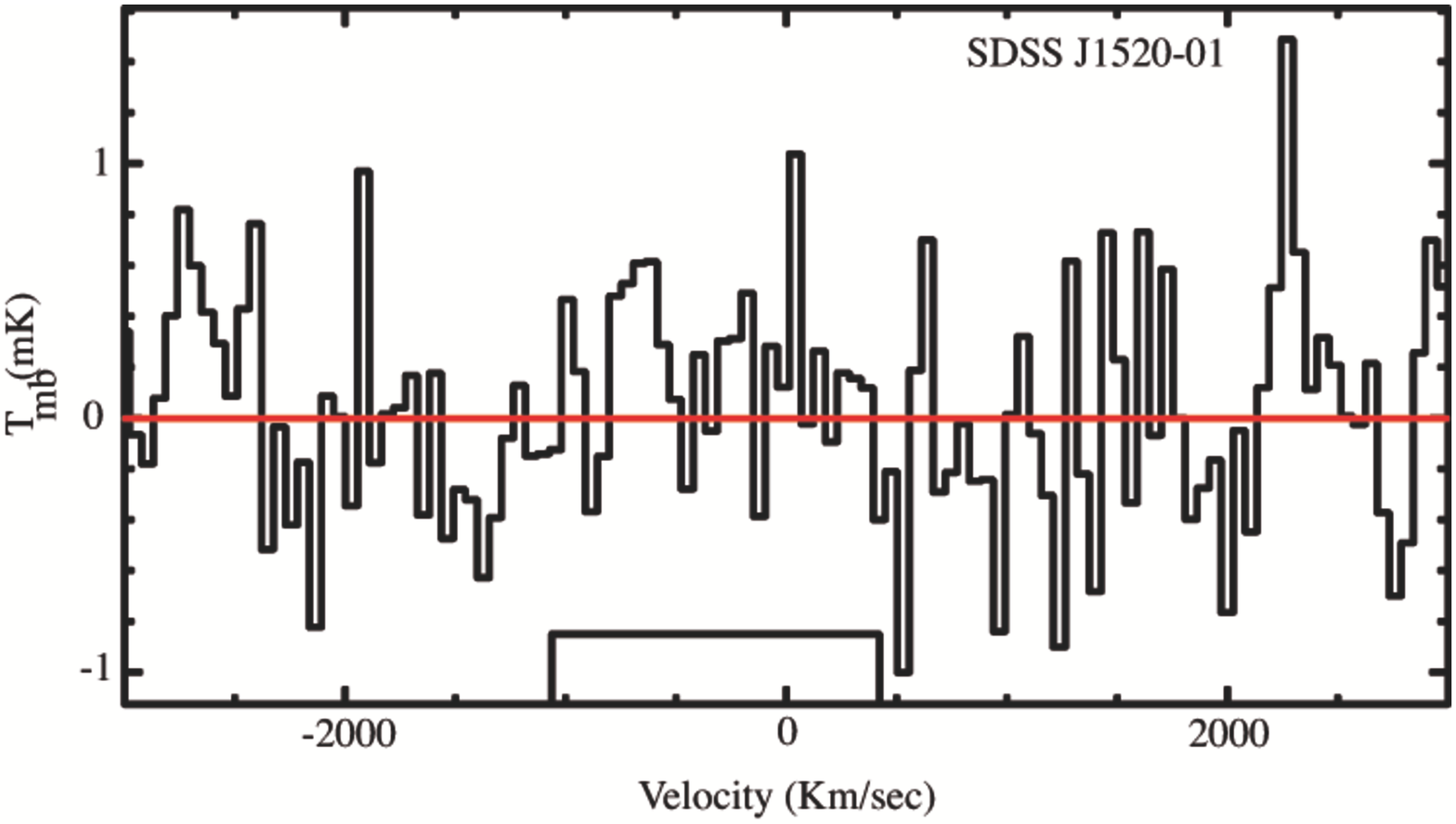}
          \hspace{-0.10in}
     \subfigure
         {\label{fig:f4}}
          \includegraphics[width=.38\textwidth, angle= 0]{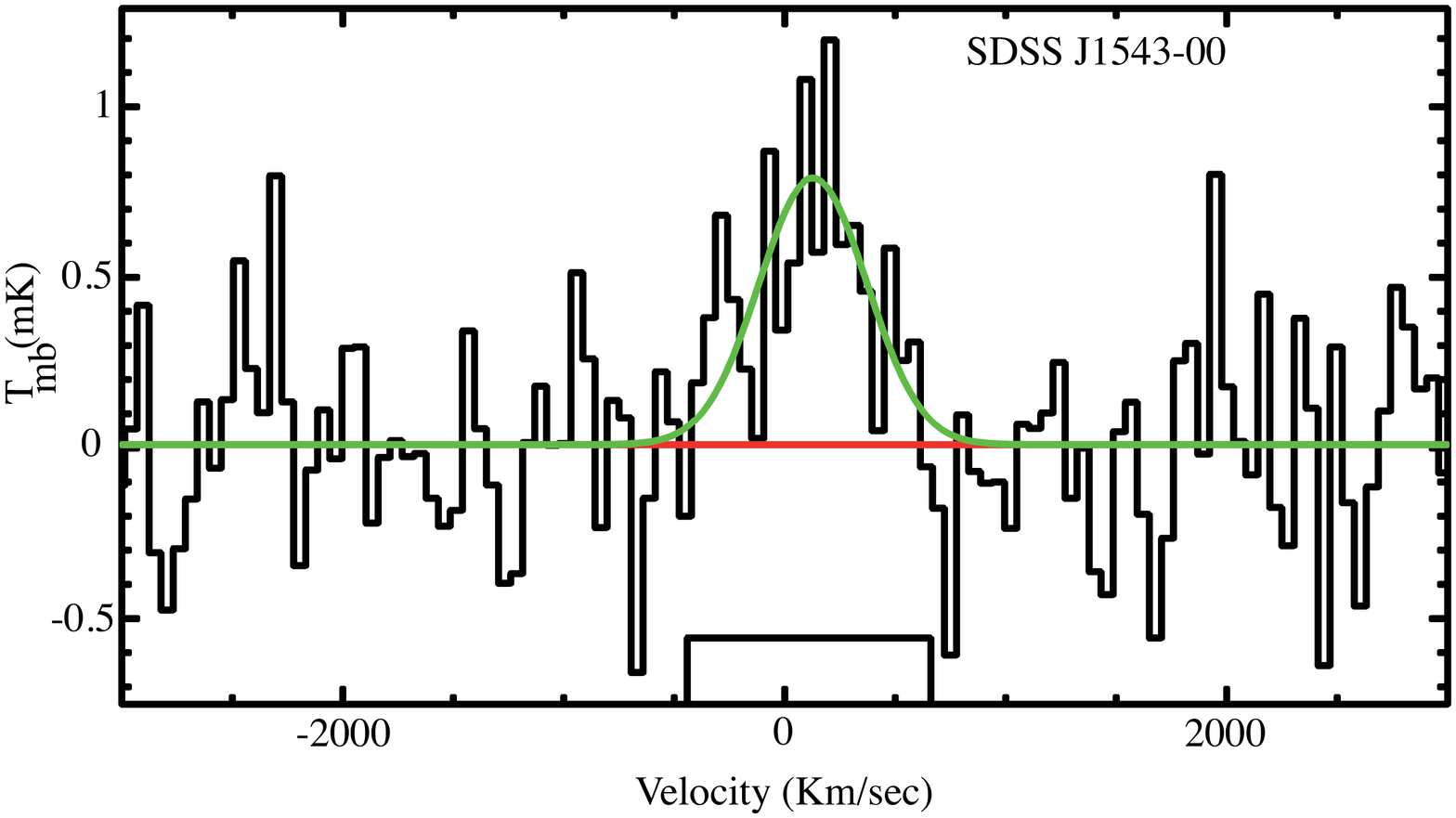}
\caption{CO(1-0) spectra of the 4 QSO2 in our sample. The zero velocity corresponds to  $z_{\rm SDSS}$.
A fit of the line profile is shown in green  for the object with detected emission.  The vertical axis shows the corrected-beam temperature. The bottom box represents the velocity window where the line is expected.}
     \label{fig:1}
\end{figure}

\section{Results}

\subsection{Infrared luminosities}

The 4 QSO2 in our sample  have an IRAS counterpart at 60 and/or 100 $\mu$m, that we have used to constrain the 
far infrared luminosities $L_{FIR}$. 
For coherence with numerous works published in the literature, we estimate $L_{FIR}$ in the 
  40-500 $\mu$m range applying  $F_{FIR}$ = 1.26$\times$10$^{-11}$  [2.58 $f_{60}$ + $f_{100}$] erg s$^{-1}$ cm$^{-2}$ and  $L_{FIR} = 4 \pi ~ D_L^2~C ~ F_{FIR}$, where we assume  $C=$1.42 (Mirabel \& Sanders 1996).  $f_{60}$ and $f_{100}$ are the IRAS flux densities at  60 and 100
 $\mu$m in Jy. Although the formula was derived for star forming galaxies at $z=$0, it is often used also for AGN and galaxies at higher $z$. 
Thus, here  we consider it  adequate
for the purpose of comparison with other works  (e.g. C11, C13, Bertram et al. 2007).

 Measurements in both the 60 and 100$\mu$m bands exist only for SDSS J0903+02, while IRAS detections are reported
only in one of the bands for the other 3 QSO2.  In such cases, we have constrained the  flux in the missing band by assuming  
typical values of $Q=f_{60 \mu m}/f_{100\mu m}$  for similar objects. There are 12 SDSS QSO2  at $z\sim$0.3-0.4
with IRAS measurements in both bands (Zakamska et al. 2004). These show $Q$ median, average and standard deviation   values 
0.26, 0.29 and 0.17 respectively.  10/12 (83\%) of these QSO2 have $Q$ in the range 0.29$\pm$0.15.
The most probable $L_{FIR}$ values and their uncertainties for SDSS J1337-01,  SDSS J1520-01 and
SDSS J1543-00 have been calculated by using  $Q$=0.29$\pm$0.15. 
   Following VM13, $L_{IR}$ was then constrained by assuming a typical
  $\xi = \frac{L_{IR}}{L_{FIR}}$=1.5. The results are shown in Table 2. The errors reflect  the uncertainty on the range of possible $Q$ values.
  The 4 QSO2 have $L_{IR}$  in the ULIRG regime ($>$10$^{12}$ L$_{\odot}$).

\subsection{$L'_{CO}$ and $M(H_2)$}

 CO(1-0) detection is confirmed in one object (SDSS J1543-00) with $L'_{CO}$=(1.2$\pm$0.2)$\times$10$^{10}$ K km s$^{-1}$ pc$^{2}$ (Table 2).
 HST images of this quasar show that it is undergoing 
a major merger event (Villar Mart\'\i n et al. 2012). Assuming  $\alpha$=0.8 M$_{\odot}$~(K km s$^{-1}$ ~pc$^2$)$^{-1}$, the implied molecular gas mass is (9.6$\pm$1.6)$\times$10$^9$ M$_{\odot}$, which is consistent with molecular gas content of other systems with similar $L_{IR}$. A 1-Gaussian fit of the line profile implies $FWHM_{CO}$=576$\pm$102 
km s$^{-1}$ with a velocity redshift relative [OIII]$\lambda$5007 of 128$\pm$47 km s$^{-1}$.
 No CO(1-0) is detected in the other 3 QSO2. Following
the same procedure described in VM13,  3$\sigma$ upper limits on the molecular gas masses are estimated to be  9.6, 4.3 and 5.4 $\times$10$^9$
  M$_{\odot}$ for SDSS J0903+02, SDSS J1337-01, SDSS J1520-01  respectively. 
    
   The location of the 4 objects in the  $L'_{CO}$ vs. $z$  and $L'_{CO}$ vs. $L_{FIR}$ and diagrams is shown in Fig.~2 as red hollow circles (see VM13 for a detailed discussion),  together with other QSO1 (blue symbols)  and QSO2 (green symbols) 
   samples (left panels) and star forming ULIRGs at $\la$1 (right panel).  The 9 ULIRG  QSO2 from C11 and C13 are represented 
with  green solid squares.  These 13   ULIRG QSO2  at $z<$1 fall on the  $L'_{CO}$ vs. $L_{FIR}$ correlation defined by other quasar samples at different $z$ and overlap with the location of star forming ULIRG. $L_{FIR}$ is likely to be dominated by the starburst contribution in those QSO2 with
total $L_{IR}<$2$\times$10$^{12}$ L$_{\odot}$. This is less certain at higher luminosities, where an increasing relative contribution of the AGN might occur (Nardini et al. 2010).  Indeed, several QSO2 and QSO1 from Combes et al. (2011, 2013) are at the lower envelope of the data distribution in the  $L'_{CO}$ vs. $L_{FIR}$ diagram.
This might suggest a poor gas content compared with star forming ULIRGs of similar $L_{FIR}$. It is however possible that this effect is a consequence of
a substantial contribution of the AGN to the  $L_{FIR}$, since all of them  have $L_{IR}>$2$\times$10$^{12}$ L$_{\odot}$, above the turning point identified by Nardini et al. (2010).

 \begin{figure*}
\includegraphics{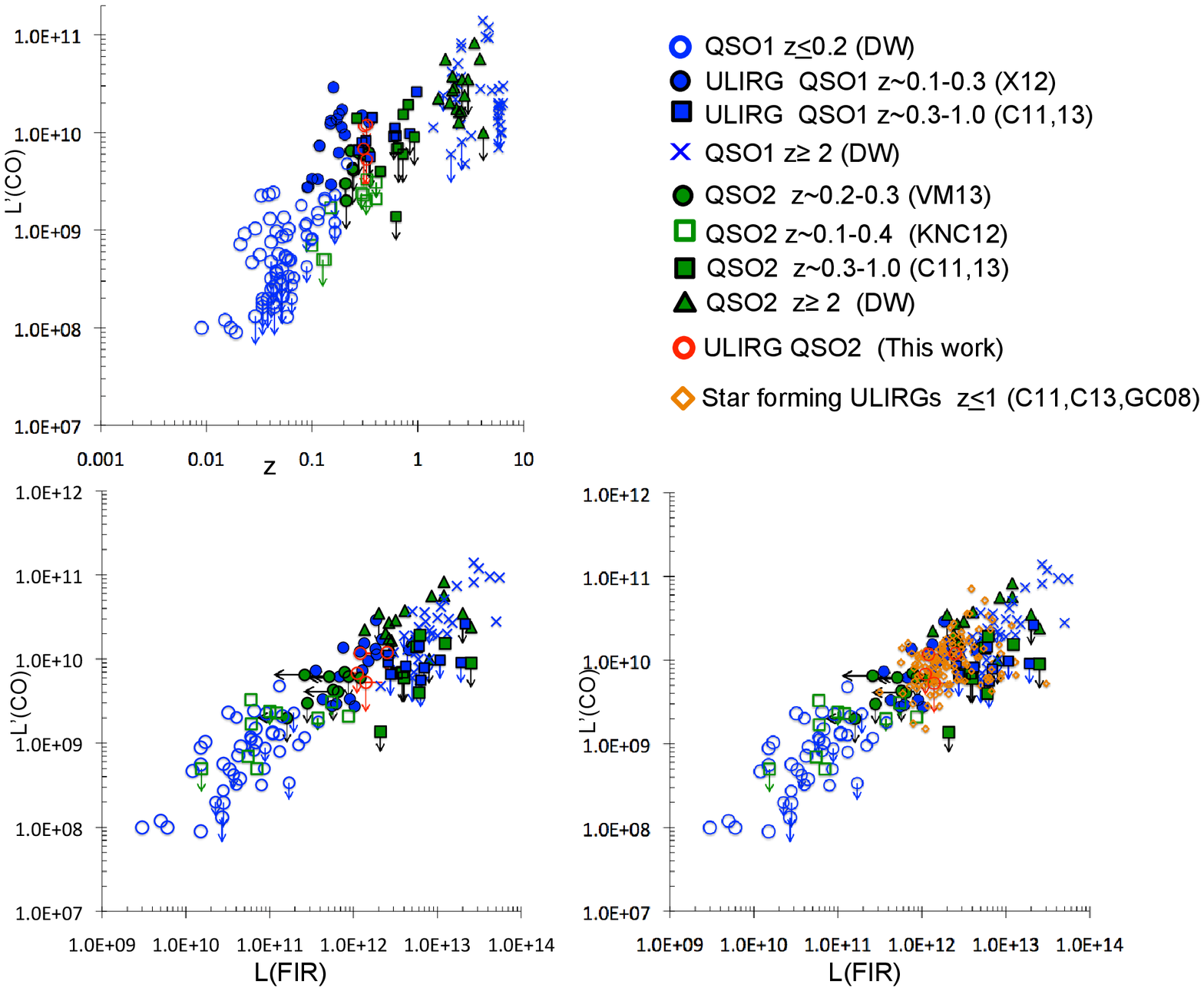}
\vspace{3.2in}
\caption{$L'_{CO}$ vs. $z$ (top) and   $L'_{CO}$ vs. $L_{FIR}$ (bottom) for  QSO1 (blue symbols) and QSO2
 (green symbols).  Our 4 ULIRG QSO2  are the red open circles. References are as follows: green solid circles: QSO2 from VM13. Open green squares: QSO2 from KNC12. Green solid squares:
QSO2 from C11 and C13. Green solid triangles: $z\ge$2 QSO2 from  different works (DW)  (Aravena et al. 2008; 
Mart\'\i nez Sansigre et al. 2009; Yan et al. 2010; Polleta et al. 2011; Lacy et al. 2011) //  Blue crosses: $z\ge$2 QSO1 from  different works (DW)  (Cox et al. 2002; Carilli et al. 2002; Walter et 
al. 2004; Krips et al. 2005; Riechers, Walter \& Carilli 2006; Gao et al. 2007; Maiolino et al. 2007; Coppin et al. 
2008; Wang et al. 2010, 2011; Riechers, Walter \& Frank 2009; Riechers et al. 2009, 2006).  Blue solid squares:
QSO1 from C11 and C13. Open blue circles: $z\le$0.2 QSO1 from different works (DW) (Evans et al. 2001, 2006; Scoville et al. 2003; Pott et al. 2006; Bertram et al. 
2007).  Filled blue circles: ULIRG QSO1 from Xia et al. 2012 (X12). Orange small diamonds: star forming ULIRGs at $z\le$1 (C11, C13; Graci\'a Carpio et al. 2008 (GC08))}
\end{figure*}

\section{Conclusions}
\label{sec:con}

$\bullet$ We present the results of CO(1-0) spectroscopic observations of 4 SDSS QSO2 at $z$ $\sim$ 0.3 
observed with the 30m IRAM telescope.  These  QSO2 have infrared luminosities in the ULIRG regime, expanding our previous work on
less infrared luminous QSO2 at $z\sim$0.3 into this regime.  The 4 QSO2 have $L_{FIR}\sim$(1-2.5)$\times$10$^{12}$ L$_{\odot}$.  We have also added 9 ULIRG QSO2 at $z\sim$0.3-0.9 from Combes et al. (2011,2013),  which have  $L_{FIR}\sim$(0.2-2.5)$\times$10$^{13}$ L$_{\odot}$.  \\
$\bullet$  CO(1-0) detection is confirmed in one of the 4 objects observed by us: SDSS J1543-01.   L$'_{CO}$ and 
$M_{H_2}$ are (1.2$\pm$0.2) $\times$10$^9$ K km s$^{-1}$ pc$^2$ and (9.6$\pm$1.6)$\times$10$^9$ 
M$_{\odot}$, respectively.  The line has $FWHM=$575$\pm$102 km $^{-1}$. No CO(1-0) emission is detected in SDSS J0903+02, SDSS J1337-01, SDSS J1520-01.
3$\sigma$ upper limits on $M(H_2)$ are 9.6, 4.3 and 5.4 $\times$10$^9$
  M$_{\odot}$ respectively.\\
$\bullet$  The $L'_{CO}$ (measurements and upper limits) of all 13 ULIRG QSO2 at $z<$1  fall in the $L'_{CO}$ vs. $L_{FIR}$ and $L'_{CO}$ vs. $z$ correlations
defined by QSO1 and QSO2 with different $z$ and $L_{IR}$. They overlap as well with  $z<$1 star forming ULIRGs. Several QSO1 and QSO2 in Combes et al. (2011, 2013) mark the lower envelope defined by the scatter of the correlation. They might be gas poor objects. Alternatively, this result might be an effect of  a significant contribution of the AGN to the $L_{FIR}$.


\begin{thebibliography}{}
\bibitem[\protect\citeauthoryear{Aravena et al.}{2008}]{ara08}
Aravena M., Bertoldi F., Schinnerer E., et al. 2008, A\&A, 4191, 173 55 
\bibitem[\protect\citeauthoryear{Bertram}{2007}]{bertram07}
Bertram T., Eckart A., Fischer S., Zuther J., Straubmeier C., Wisotzki L., Krips M., 2007, A\&A, 470, 571 
\bibitem[\protect\citeauthoryear{Carilli et al.}{2002}]{ca02}
Carilli C., Kohno K., Kawabe R., et al. 2002, AJ, 123, 1838
\bibitem[\protect\citeauthoryear{Combes et al.}{2011}]{co11}
Combes, F., Garc\'\i a-Burillo, S., Braine, J., Schinnerer, E., Walter, F., Colina, L., 2011, A\&A, 528A, 124
\bibitem[\protect\citeauthoryear{Combes et al.}{2013}]{co13}
Combes, F., Garc\'\i a-Burillo, S., Braine, J., Schinnerer, E., Walter, F., \& Colina, L., 2013, A\&A, 550A, 41
\bibitem[\protect\citeauthoryear{Coppin et al.}{2008}]{coppin08}
Coppin K., Swinbank A. M., Neri R., et al. 2008, MNRAS, 389, 45 
\bibitem[\protect\citeauthoryear{Cox et al.}{2002}]{cox02}
Cox P., Omont A., Djorgovski S. G., et al. 2002, A\&A, 387, 406
\bibitem[\protect\citeauthoryear{Downes \& Solomon}{1998}]{dow98}
Downes, D. \& Solomon, P. M., 1998, ApJ, 507, 615
\bibitem[\protect\citeauthoryear{Evans et al.}{2001}]{e01}
Evans A. S., Frayer D. T., Surace J. A., Sanders D. B., 2001, AJ, 121, 1893 
\bibitem[\protect\citeauthoryear{Evans et al.}{2006}]{e06}
Evans A. S., Solomon P. M., Tacconi L., Vavilkin T., Downes D., 2006, AJ, 132, 2398 
\bibitem[\protect\citeauthoryear{Gao et al.}{2007}]{gao07}
Gao Y., Carilli C., Solomon P., Vanden Bout P., 2007, ApJ, 660, 93
\bibitem[\protect\citeauthoryear{Graci\'a Carpio et al.}{2008}]{Gra08}
Graci\'a Carpio J., Garc\'\i a-Burillo S., Planesas P., Fuente A., Usero A., 2008,
A\&A, 479, 703
\bibitem[\protect\citeauthoryear{Guilloteau \& Forveille}{1989}]{gui89}
Guilloteau S., \& Forveille T., 1989, 
\\ \textit{Grenoble Image and Line Data
Analysis System (GILDAS)}, 
\\ IRAM, http://www.iram.fr/IRAMFR/GILDAS
\bibitem[\protect\citeauthoryear{Krips et al.}{2005}]{krips05} 
Krips M., Eckart A., Neri R. et al. 2005,  A\&A, 439, 75
\bibitem[\protect\citeauthoryear{KNC12}{}]{knc12} 
Krips M., Neri R., Cox P., 2102, ApJ, 753, 135 (KNC12)
\bibitem[\protect\citeauthoryear{Lacy et al.}{2007}]{lacy07} 
Lacy, M., Sajina, A., Petric, A. O., et al., 2007, ApJ, 669L, 61
\bibitem[\protect\citeauthoryear{Lacy et al.}{2011}]{lacy011} 
Lacy M., Petric A., Martnez-Sansigre A., Ridgway S., Sajina A., Urrutia T., Farrah D., 2011, AJ, 142, 196 24 
\bibitem[\protect\citeauthoryear{Maiolino et al.}{2007}]{mai07}
Maiolino R., Neri R., Beelen A., et al. 2007, A\&A, 472, L33
\bibitem[\protect\citeauthoryear{Mart\'\i nez-Sansigre et al.}{2006}]{mar06}
Mart\'\i nez-Sansigre, A., Rawlings, S., Garn, T., et al. 2006, MNRAS, 373L, 80
\bibitem[\protect\citeauthoryear{Mar\'\i nez Sansigre et al.}{2009}]{sans09} 
Mart\'\i nez Sansigre A., Karim A., Schinnerer E. et al., 2009, ApJ, 706, 184
\bibitem[\protect\citeauthoryear{Nardini et al.}{2010}]{nardini10}
Nardini, E., Risaliti, G., Watabe, Y., Salvati, M., \& Sani, E., 2010, MNRAS, 405, 2505
\bibitem[\protect\citeauthoryear{Papadopoulos et al.}{2012}]{pap12}
Papadopoulos P., van der Werf P., Xilouris E., Isaak K., Gao Y., 2012, ApJ, 751, 10
\bibitem[\protect\citeauthoryear{Pier \& Krolik}{1992}]{pk92}
Pier, E. A., \& Krolik, J. H., 1992, ApJ, 401, 99 
\bibitem[\protect\citeauthoryear{Polleta et al.}{2011}]{pol11} 
Polletta  M., Nesvadba N., Neri R., Omont A., Berta S., Bergeron J., 2011, A\&A, 533, 20 
\bibitem[\protect\citeauthoryear{Pott et al.}{2006}]{pott06}
Pott, J.-U., Eckart, A., Krips, M., Tacconi-Garman, L. E., \& Lindt, E., 2006, A\&A, 456, 505
\bibitem[\protect\citeauthoryear{Riechers, Walter \& Carilli}{2006}]{rie06}
Riechers D., Walter F., Carilli C. et al. 2006, ApJ, 650, 604
\bibitem[\protect\citeauthoryear{Riechers et al.}{2009}]{riechers09}
Riechers D., Walter F., Carilli C., Lewis G., 2009, ApJ, 690, 485
\bibitem[\protect\citeauthoryear{Riechers, Walter \& Frank}{2009}]{rie09}
Riechers D., Walter F., Frank B., 2009, ApJ, 703, 1338
\bibitem[\protect\citeauthoryear{Sanders \& Mirabel}{1996}]{sanders96}
Sanders, D. B., \& Mirabel, I. F., 1996, ARA\&A, 34, 749
\bibitem[\protect\citeauthoryear{Scoville et al.}{2003}]{scoville03}
Scoville N., Frayer D., Schinnerer E., Christopher M., 2003, ApJ, 585, L105 22 
\bibitem[\protect\citeauthoryear{Szokoly et al.}{2004}]{szokoly04} 
Szokoly, G. P.; Bergeron, J.; Hasinger, G., et al., 2004, ApJS, 155, 271
\bibitem[\protect\citeauthoryear{VM12}{}]{vm12} 
Villar Mart\'\i n M.,  Cabrera Lavers A., Bessiere P., Tadhunter C., Rose M, de Breuck C.,  2012, MNRAS, 423, 80 [VM13]
\bibitem[\protect\citeauthoryear{VM13}{}]{vm13} 
Villar Mart\'\i n M., Rodr\'\i guez M., Drouart G. et al. 2013, MNRAS, 434, 978
\bibitem[\protect\citeauthoryear{Walter et al.}{2004}]{wa04}
Walter F., Carilli C., Bertoldi F. et al. 2004, ApJ, 615, L17
\bibitem[\protect\citeauthoryear{Wang et al.}{2010}]{wang10}
Wang R., Carilli C., Neri R., et al. 2010, ApJ, 714, 699 
\bibitem[\protect\citeauthoryear{Wang et al.}{2011}]{wang11}
Wang R., Wagg J., Carilli C. L. et al. 2011, AJ, 142, 101 
\bibitem[\protect\citeauthoryear{Wu et al.}{2010}]{wu10}
Wu Y. et al., 2010, ApJ, 723, 895
\bibitem[\protect\citeauthoryear{Xia et al.}{2012}]{xia12}
Xia, X. Y., Gao, Y., Hao, C.-N., et al. 2012, ApJ, 750, 92
\bibitem[\protect\citeauthoryear{Yan et al.}{2010}]{yan10}
Yan L., Tacconi L., Fiolet N., Sajina A., Omont A., Lutz D., Zamojski M., Neri R., Cox P., Dasyra K., 2010, ApJ, 714, 100 55 
\bibitem[\protect\citeauthoryear{York et al.}{2000}]{york00} 
York D. G., Adelman J., Anderson S. et al., 2000, AJ, 120, 1579
\bibitem[\protect\citeauthoryear{Zakamska et al.}{2003}]{zak03} 
Zakamska N., Strauss M., Krolik J. et al. 2003, AJ, 126, 2125
\bibitem[\protect\citeauthoryear{Zakamska et al.}{2004}]{zak04} 
Zakamska N., Strauss M., Heckman, T. et al. 2004, AJ, 128, 1002

\end{thebibliography}
\end{document}